# Spatial Filtering of Sound Beams by Sonic Crystals

R. Picó[1], V. J. Sánchez-Morcillo[1], I. Pérez-Arjona[1] and K. Staliunas[2]

[1]Instituto de Investigación para la Gestión Integrada de Zonas Costeras, Universidad Politécnica de Valencia, C/. Paranimf 1, 46730 Grao de Gandía (Valencia), España

[2]ICREA, Departament de Fisica i Enginyeria Nuclear, Universitat Politècnica de Catalunya, Colom 11, E-08222 Terrassa (Barcelona) España

***Abstract:*** *We propose and numerically demonstrate an efficient cleaning of spatial structure (spatial filtering) of sound beams by propagating them through at least two-dimensional sonic crystals, i.e. through acoustic structures periodically modulated in longitudinal and in transversal direction with respect to the sound propagation direction. We show the spatial filtering in two configurations: with- and without the angular band-gap. We also show that besides the spatial filtering the beams can be additionally focalized at a particular distance behind the sonic crystal in both configurations.*

The acoustic beams emitted by typical sound sources, as for example a piezoelectric transducer, are often of a low spatial quality. Roughly speaking, the spatial quality is related to inhomogeneities in the beam intensity and phase profile, and is a measure of how tight the beam can be focused: the higher is the beam quality, the narrower is the beam waist in the focal region. It is commonly accepted that the beams of the highest spatial quality are these of gaussian profile. The low spatial quality beams can be focused worse in comparison with an "ideal" gaussian beam, therefore their applications (e.g. in tomography, microscopy, imaging, sonar, etc) are not optimal. Furthermore, in the case of unfocused beams, a higher spatial quality results in a lower diffraction divergence, the gaussian beam being again an example of the beam with minimum divergence. Some definitions and parameters to quantify the spatial quality of the beams can be found in [1].

One possibility to obtain the beams of higher spatial quality is to optimise the transducers used to emit the beams. For instance, the design of gaussian acoustic transducers was proposed in [2]. The other possibility, on which we concentrate here, is to "clean" a noisy beam (or more generally, to reduce the width of its spatial spectrum) with a spatial filter. This technique, widely used in optics [3], has to our knowledge no analogue for acoustic beams. We propose here an efficient method for the spatial filtering or cleaning of the spatial structure of sound beams by propagating them through a structure periodic in space, with the period of the wavelength scale - the so called Sonic Crystal (SC).

The spatial filtering mechanism in SCs is easily understandable in the case when the propagation eigenmodes (the Bloch modes) of the sound field display angular band-gaps. As the frequency band-gaps in the temporal dispersion relation $\omega(k)$ of periodic materials can be utilized to manipulate (to filter out) particular regions of the temporal spectrum [4], analogously the angular band-gaps should allow manipulating the angular, or the spatial spectrum. We recall that a sound beam in homogeneous media can be decomposed into plane waves propagating at different angles, and the sound beam in modulated media can be decomposed into Bloch modes propagating at different angles. The spatial and the angular distributions of the beam are related via the Fourier transform: the wider is the spatial beam width, the narrower is its spatial spectrum. The parameters of the crystal (mainly the longitudinal and the transverse periods, but also the filing ratio) can be varied in order to obtain a transmission curve as illustrated in Fig.1.a. In order to build an efficient spatial filter one aims to achieve high transparency for the axial and small-angle components, and low transparency (high reflectance) for the particular angular components which are to be filtered out.

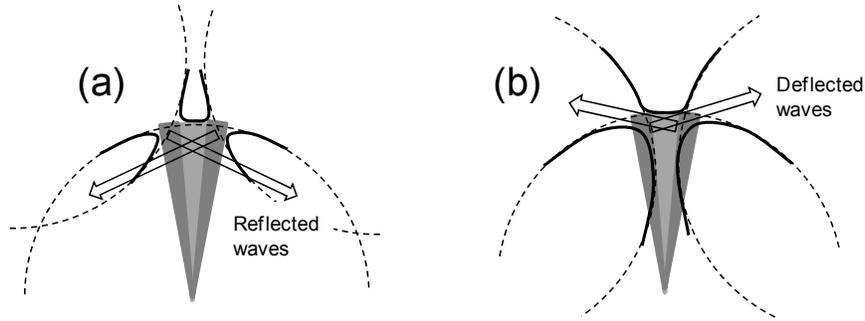

*Fig.1. Illustration of the spatial filtering in the case of a SC, in the spatial Fourier plane ($k_x$,$k_y$): (a) with angular bandgaps, and (b) without angular bandgaps. Spatial filtering occurs in both cases around the angles where the dispersion curves of the harmonic components for a particular frequency (dotted circles) mutually cross. The central part of the upper dispersion curve is responsible for the focalization of the filtered beam behind the SC. The spatial spectrum (far field) of the initial beam consisting of the central (regular) part, and the wings (the part to be removed, i.e. reflected in (a) or deflected in (b)) is illustrated with bright and dark triangles. The central dashed circle indicates the spatial dispersion of the homogeneous wave in homogeneous material, the other circles indicates the dispersion of the harmonic components in SC.*

Although such mechanism of spatial filtering is intuitively well understandable, however, to the best of our knowledge, up to now has been nowhere clearly demonstrated, neither experimental nor numerically. We show in this article the feasibility of the filtering mechanism by numerical simulation of the propagation of a sound beam through a SC presenting angular bandgaps for a given incident beam. However, we show that there exists another, less evident mechanism of spatial filtering in SCs, in the configuration displaying no angular band-gaps. That latter mechanism of filtering in a gapless configuration is somewhat related to that recently proposed in optics in [5]. The gapless filtering is based on the idea that particular shapes of the spatial dispersion curve (in particular the strongly tilted and strongly curved segments of the spatial dispersion curve) can deflect certain angular components of the beam. Fig.1.b illustrates the latter mechanism, which does not require the presence of the angular band-gap. Differently from the first mechanism the angular components are not reflected at the entrance of sound wave into SC, but instead deflected into the first diffraction maxima.

The angular bandgaps occur if the dispersion circles of the plane-wave components cross at angles larger than 90 degrees (the dashed circles in Fig.1.a). The corresponding geometry of the SCs is that the longitudinal modulation period (lattice constant) $d_\parallel$ is less than the transversal one $d_\perp$. Considering a SC of a rhombic configuration it means that the sound beam propagates along the shorter diagonals of the rhombic cells. Then the plane wave components of the beam with wavevectors lying in the area around the crossing of the dispersion curves are scattered into the backward direction. If the dispersion circles cross at angles smaller than 90 degrees then the corresponding plane waves are scattered at diffraction angles, however to the forward direction (perpendicular to the segments of the dashed circles in Fig. 1.b). The angular gaps do not appear in this case, nevertheless the particular angular components of the beam can be removed, and the incident beam can be spatially filtered. The latter situation occurs for sound beams propagating along the large diagonals of the rhombic SC structure.

In the present paper we report the numerical evidence of both mechanisms of spatial filtering. We engineer the SC structures by calculating the two-dimensional band diagrams (iso-frequency curves, IFC) in both cases, by using the plane wave expansion (PWE) method, and identify the angular band-gaps and the strongly curved segments of the spatial dispersion curve. We note, that we could find the SC structure where the both filtering mechanisms can be realised in two orthogonal directions of the same structure (however for different frequencies). The spatial filtering is demonstrated in both cases by realistic numerical calculations of the full model (without assuming the paraxial approximation) using the finite element method [6].

However, besides the spatial filtering of the sound beam after its passage through the SC, which is the basic phenomenon reported in the present paper, we observe an additional side-effect of focalisation of the filtered beams at some distance behind the SC. The latter phenomenon is due to a nonzero curvature of the central segment of the dispersion curve, and was already discussed in [6]. The focalisation behind the SC occurs when the curvature of the central segment of dispersion curve is positive, as then the SC presents an anomalous diffraction (i.e. the diffraction coefficient has negative sign). Therefore the SC has a double effect on the incident sound beam: it predominantly acts as a spatial filter, and additionally it also acts as a focusing lens.

As a theoretical model for the analysis of the propagation of sound through the SC we consider the inhomogeneous wave equation for the is the scalar pressure field $p(\mathbf{r},t)$,

$$\frac{1}{B(\mathbf{r})}\frac{\partial^2 p(\mathbf{r},t)}{\partial t^2} + \nabla\left(\frac{1}{\rho(\mathbf{r})}\nabla p(\mathbf{r},t)\right) = 0 \qquad (1)$$

where the functions $B(\mathbf{r})$ and $\rho(\mathbf{r})$ are, respectively, the spatially dependent (periodic) bulk modulus and density of the SC. The design of the structure showing the different filtering mechanisms (presenting or not angular band-gaps depending on the direction of incidence) is done by calculating the spatial dispersion curves, assuming a monochromatic wave in the form $p(\mathbf{r},t) = p(\mathbf{r})\exp(i\omega t)$, and finding the eigenvalues $\omega(k)$ from Eq. (1) using the PWE method [6-8]. The results are summarised in Fig. 2 for both configurations, with and without the angular bandgaps.

The sonic crystal used in the simulations consists of a periodic array of steel cylinders with radius $r = 0.8$ mm immersed in water. The material parameters are $\rho_h = 10^3$ Kg m$^{-3}$ and $B_h = 2.2 \cdot 10^9$ N m$^{-2}$ for the host medium (water) and $\rho_s = 7.8 \cdot 10^3$ Kg m$^{-3}$, $B_s = 160 \cdot 10^9$ N m$^{-2}$ for steel, with corresponding sound velocities $c_h = 1483$ m s$^{-1}$ and $c_s = 4530$ m s$^{-1}$. In such a crystal, the frequencies of interest correspond to the ultrasonic regime.

A SC with rhombic symmetry is considered, with the length of the short and long diagonals given by $d_l = 9.25$ and $d_s = 6.2$ mm. The corresponding IFC are shown in Fig. 2, for particular values of the frequencies. In Fig. 2(a) the frequencies $f = 140$ kHz (grey lines), $f = 150$ kHz (black lines) and $f = 160$ kHz (dashed lines) are shown. Figure 2(b) corresponds to $d_l = 6.2$ and $d_s = 9.25$ mm, which can be understood as the crystal shown in Fig. 2(a) but rotated by 90º. The shown frequencies are $f = 200$ kHz (black lines) and $f = 220$ kHz (grey lines) in this case. In Fig. (2) the values $f = 200$ and 140 kHz were chosen because there exist a flat region in the corresponding IFC, for propagation along the short and the long diagonals of the SC respectively. These frequencies correspond to the self-collimation frequencies [8,9] and the beam, apart from being spatially filtered is not expected to experience any propagation effects through the SC. For frequencies larger than the self-collimation frequencies, the dispersion curve obtains a positive curvature, which allows to expect a focalisation, or at least a reduction of the divergence behind the SC.

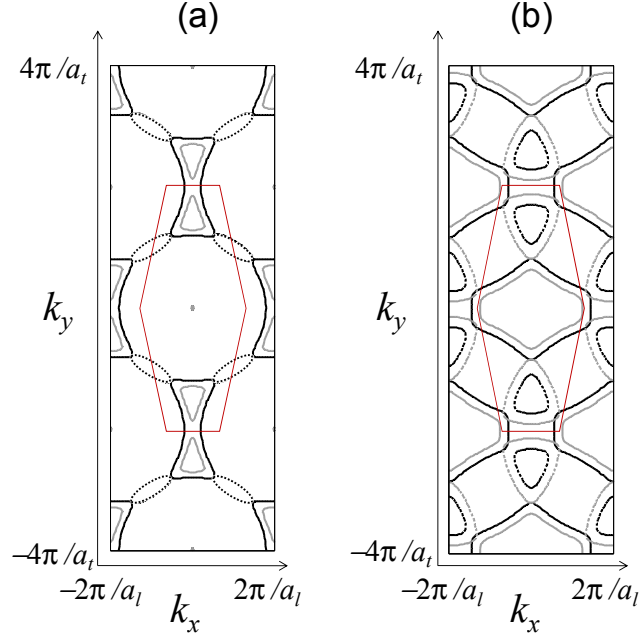

*Fig.2.* (Color online) Iso-frequency contours of the Bloch modes in a SCs with $d_\parallel = 6.2$ mm and $d_\perp = 9.25$ mm (a), and with $d_\parallel = 9.25$ mm and $d_\perp = 6.2$ mm (b), evaluated for different frequencies. In (a) the curves correspond to f=140 kHz (black, solid line), 150 kHz (grey line) and 160 kHz (black, dashed line), while in (b) the curves corresponding to f=200 kHz (black lines) and f=220 kHz (grey lines) are shown. A beam propagating along $k_y$ in the case (a) experience angular band-gaps while a beam along $k_x$ in the case (b) does not. The flat regions of IFC correspond to self-collimating regimes. The coloured polygon delimits the first Brillouin zone, whose limits are $k_x = \pm\pi/a_l$ and $k_y = \pm 2\pi/a_t$.

Once the optimal structure and the frequencies are determined from the PWE calculations, we simulate the sound beam propagation in such structure by solving Eq. (1) using the finite element method [6]. In order to check the value of the optimum frequency obtained from the PWE we also calculate the dependence of the beam profile in the far-field on the frequency [Figs. 3(a) and (b)]. The far field is obtained by evaluating the Fourier transform of the near field, corresponding to the beam profile at the exit of the SC. Figure 3(a) evidences that in the configuration presenting angular band-gaps, the band-gap appears also in frequency domain. Increasing the frequency of the sound wave the central angles of the two symmetrically placed angular bandgaps move closer one to another. The temporal bandgaps appear when the two angular bandgaps join at zero angle (on sound propagation axis). In the angular gapless configuration no bandgaps appear in the frequency domain [Fig. 3(b)].

Figures 3(a) and (b) also show that the angular distribution of the field suffers a significant narrowing in both cases. The filtering angles decrease with frequency, as can be expected from the band diagrams. The character of the narrowing of the spectra depends on the frequency and on the structural parameters of the SC. For a concrete application (to selectively remove a part of the spatial spectrum) the parameters must be chosen in such a way that the undesired spatial frequencies lie in the area of filtering. In any case a significant narrowing of the angular distribution can be obtained in comparison with the initial distribution, as Figs. 3(c) and (d) show.

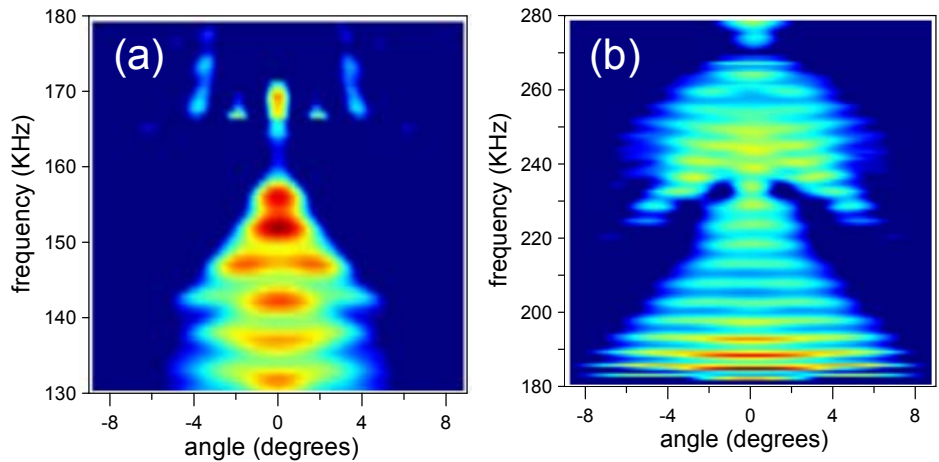

*Fig3. (Color online) The far field (angular) distribution of the acoustic intensity depending on frequency, in the configuration with angular band-gap (a) and in the gapless configuration (b). The parameters are as in Fig. 2. The angle is given in degrees.*

The beams propagating inside and behind the SC in several characteristic cases are shown in Figs. 4 and 5. The source emits a beam with a Gaussian profile, which propagates through a finite crystal with length $L$=125 mm. Again, as expected, for frequencies corresponding to the self-collimation frequency no focalisation occurs in addition to the broadening of the beam due to spatial filtering [Figs. 4(a) and 5(a)]. However for the frequencies higher than those of self-collimation, the additional beam focalization is visible [Figs. 4(b) and 5(b)].

Figures 4 and 5 also indicate the character of scattered radiation in both cases. While in the configuration presenting angular bandgaps (Fig. 4) the filtered radiation is reflected back (at some angles to the axis) at the very entrance of the SC, in the gapless configuration (Fig. 5) the filtered radiation is deflected, however propagates in forward direction. The explanation of such behaviour was discussed schematically if Fig. 1.

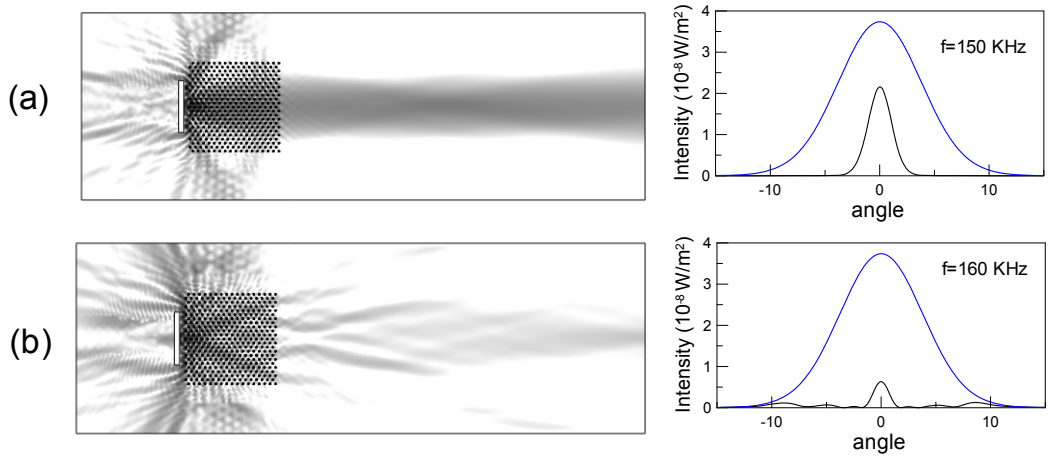

*Fig.4.* (Color online) Beam propagation in- and behind the SC for the configuration with the angular bandgaps, corresponding to the case in Fig. 2(a) with the beam propagating along $k_y$, for the self-collimating frequency f=150 kHz (a) and for higher frequency corresponding to negative diffraction (and focalization), f=160 kHz (b). On the right the spatial spectra (blue curves) in both cases are shown as compared with the angular beam profiles at the initial beam (at the entrance of the SC, black curves). Note the presence of the reflected waves, corresponding to the Bloch modes lying on an angular gap.

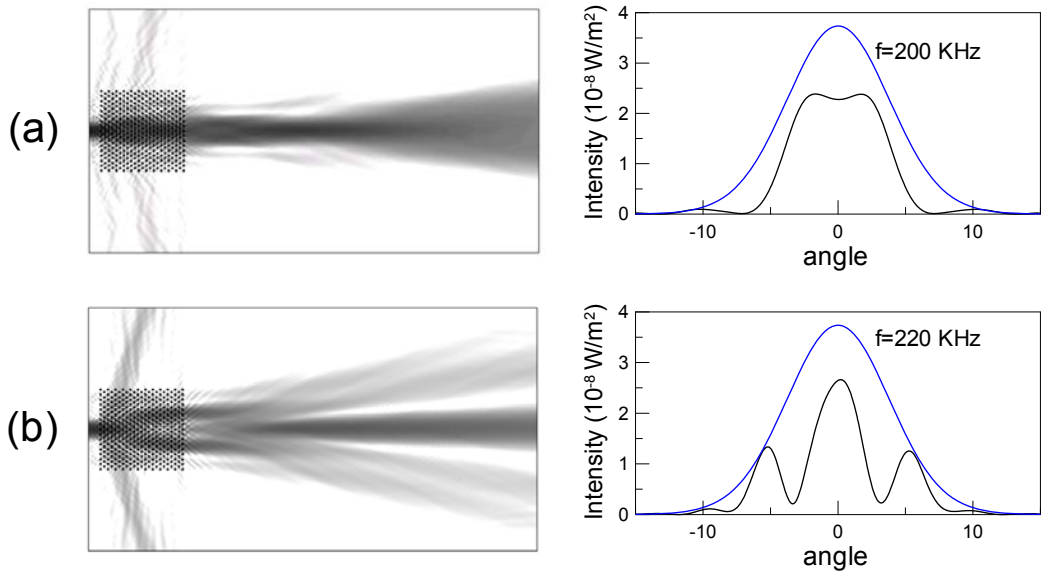

*Fig.5.* (Color online) The same as in Fig.4, but for the gapless configuration, corresponding to the case in Fig. 2(b) with the input beam along $k_x$.

Summarizing, we propose an efficient method for the modification of the spatial spectrum of a sound beam (spatial filtering) by propagating it through SCs with particular transverse and longitudinal lattice periods. We prove the proposed effect on a concrete example of 2D sonic crystal (cylindrical scatterers) of rhombic symmetry by means of a numerical integration of the sound propagation model Eq. (1). From the point of view of applications, a 3D filtering device acting on the whole transverse plane would be most convenient. However, the effects reported for the 2D systems are expected to occur in 3D. We note that, as shown in [9], there are no substantial differences concerning the optimal frequencies and crystal parameters between 2D and 3D cases.

We have also shown that the spatial filtering in SCs can be combined with the effect of sound focalization behind the SCs. In fact both effects are mutually related. This fact is very promising for the applications, as allows combining two functionalities into one.

In the article we considered the manipulation of the beams of broad spatial spectra, however not the random (noisy) beam. The bad spatial quality in the calculated examples is due to the use of a realistic transducer, which excites the beams with particular (rectangular) spatial profile. The presented analysis is valid also for random beams too, e.g. for those which appear after propagation in a turbulent media. The propagation through the SC is linear (we consider small amplitudes) therefore the beam is a linear composition of the components of its spatial spectrum. Each angular component can be manipulated independently by the SC. In this way the discussed spatial filtering is completely independent on the nature of the broadening of the spectrum of the beam.

The work was financially supported by Spanish Ministerio de Ciencia e Innovación and by European Union FEDER through projects FIS2008-06024-C02-02 and -03

*References:*

[1] A. E. Siegman, "Defining, measuring, and optimizing laser beam quality", Proc. SPIE 1868, 2 (1993);

[2] D. Huang and M.A. Breazeale, "An ultrasonic Gaussian transducer and its diffraction field: theory and practice", IEEE Transactions on Ultrasonics, Ferroelectrics and Frequency Control, **53**, 1018-1027 (2006)

[3] In optics the filtering is performed by propagating a tightly focussed beam through a diaphragm, which blocks the higher spatial (angullar) components. The metod proposed in the present article (filtering by periodic material, i.e. photonic crystal) has never been demonstrated in optics.

[4] A. Khelif, B. Aoubiza, S. Mohammadi, A. Adibi and V. Laude, Phys. Rev. E **74**, 046610 (2006).

[5] K. Staliunas and V. J. Sánchez-Morcillo, Phys. Rev. A **79**, 053807 (2009)

[6] For the numerical simulations the standard computing package Comsol 3.5 was used.

[7] V.J. Sánchez-Morcillo, K. Staliunas, V. Espinosa, I. Pérez-Arjona, J. Redondo and E. Soliveres, Phys. Rev. B **80**, 134303 (2009).

[8] I. Pérez-Arjona, V.J. Sánchez-Morcillo, J. Redondo, V. Espinosa and K. Staliunas, Phys. Rev. B **75**, 014304 (2007)

[9] V. Espinosa, V.J. Sánchez-Morcillo, K. Staliunas, I. Pérez-Arjona and J. Redondo, Phys. Rev. B **76**, 140302(R) (2007)

[10] E. Soliveres, V. Espinosa, I. Pérez-Arjona, V.J. Sánchez-Morcillo, and K.Staliunas, Appl. Phys. Lett. **94**, 164101 (2009).